\documentclass{acm_proc_article-sp}
\usepackage{graphicx}
\usepackage{array}
\usepackage{amsmath}
\begin{document}

\title{A Semantic-Based Approach for Detecting \\ and Decomposing God Classes}
\numberofauthors{4}

\author{
% 1st. author
\alignauthor
Jun H. Lee\\
       \affaddr{Sogang University}\\
       \affaddr{Seoul, S. Korea}\\
       \email{zuna@sogang.ac.kr}
% 2nd. author
\alignauthor
Donghun Lee\\
       \affaddr{Sogang University}\\
       \affaddr{Seoul, S. Korea}\\
       \email{freedust@sogang.ac.kr}
% 3rd. author
\alignauthor Dae-Kyoo Kim\\
       \affaddr{Oakland University}\\
       \affaddr{MI, USA}\\
       \email{kim2@oakland.edu}
\and  % use '\and' if you need 'another row' of author names
% 4th. author
\alignauthor
Sooyong Park\\
       \affaddr{Sogang University}\\
       \affaddr{Seoul, S. Korea}\\
       \email{sypark@sogang.ac.kr}
}
%\date{30 July 1999}
\maketitle

%===========================================================================
\begin{abstract}
Cohesion is a core design quality that has a great impact on
posterior development and maintenance. 
By the nature of software, the cohesion of a system is
diminished as the system evolves. God classes are 
code defects resulting from software evolution, 
having heterogeneous responsibilities highly coupled 
with other classes and often large in size, which
makes it difficult to maintain the system. 
The existing work on identifying and decomposing God classes 
heavily relies on internal class information to identify 
God classes and responsibilities.
However, in object-oriented systems, responsibilities should 
be analyzed with respect to not only internal class
information, but also method interactions. 
In this paper, we present a novel approach 
for detecting God classes and decomposing their responsibilities
based on the semantics of methods and method interactions. 
We evaluate the approach using JMeter v2.5.1 and 
the results are promising. 

\end{abstract}
%===========================================================================

\section{Introduction}
\label{sec:introduction}
%=======================================================================================
Object-oriented development (OOD) is responsibility-driven.
A class is assigned a single responsibility to carry out its intended purpose,
having high cohesion. 
However, by the nature of software evolution, a single responsibility
is often diminished with other responsibilities mixed by changes, which
decreases the cohesion of the class and further that of the system as a whole.
Such a class is desired to be restructured.

Software evolution is often ad-hoc, which makes it difficult to
identify classes needing refactoring. Class size is often used as
an initial screening~\cite{Rapu2004}. However, even though
class size is small, the class might still need refactoring
if it involves multiple responsibilities.
High fan-in and fan-out is another symptom of God 
classes~\cite{Chatzigeorgiou2004}. However, that is not always the case, 
for example facade or proxy classes whose the main responsibility is delegation. 
Another symptom is high complexity of methods~\cite{Chidamber1994}.
A sorting class often involves complex methods, but is generally not large. 
As such, identifying refactoring needs can be subjective depending on
the type of the system and developer's experience and 
requires techniques that enable systematic detection 
in consideration of various aspects.

God classes (also known as Large classes or Blobs)
are code defects resulting from software evolution,
having diverse responsibilities highly coupled
with other classes and often large in size, which
makes it difficult to maintain the system.
Thus, the more God classes exist, the lower cohesion is.
God classes might be inherent from design during development, which 
is a design defect~\cite{Marinescu2004}.

There is some work on decomposing God classes (also known as
class extraction or refactoring).
The general approach of the existing work is using
internal class information such as attribute-method
relationships and internal method calls
to identify class responsibilities \cite{Simon2001,Chatzigeorgiou2004,Joshi2009,Cassell2009,Fokaefs2009}.
However, in object-oriented systems, identifying responsibilities
solely based on internal class information is very limited
without considering interaction behaviors.
More recent work makes use of semantic similarity of methods
captured in in-line comments and identifier names,
assuming that the necessary information is
sufficiently available~\cite{Lucia2008,Bavota2010,Bavota2011397}.

A key in decomposing responsibilities is to derive precise semantics of methods,
so that homogeneous methods can be identified and grouped together into a separate class.
In this paper, we present a semantic-based approach for detecting God classes and
identifying their responsibilities based on semantic similarity of methods.
Semantic similarity of methods
is measured based on 1) inter-class interactions of methods, 2) intra-class
interactions of methods, and 3) types of class relationships.
We adopt the taxonomy by Resnik \cite{Resnik1995} for analyzing inter-class interactions
of methods. The results of the taxonomy are refined by considering intra-class
interactions of methods and further refined using types of class relationships.
The refined results are used for detecting God classes using weighted graphs
and decomposing their responsibilities.
We evaluate the presented approach using JMeter v2.5.1, a widely used open source application
for load testing and measuring server performance and the results are promising. 
We design the approach to support its use at both design level and code level.

The remainder of the paper is organized as follows.
Section~\ref{sec:relatedwork} discusses an overview of related work.
Section~\ref{sec:overview} gives an overview of the presented approach.
Section~\ref{sec:csm} describes a structural taxonomy for measuring and refining semantic similarity of methods.
Section~\ref{sec:badsmell} presents detecting God classes and identifying and decomposing their responsibilities.
Section~\ref{sec:casestudy} evaluates the presented approach using JMeter v.2.5.1
and the paper is concluded in Section~\ref{sec:conclusion}.

%=======================================================================================
\section{Related Work}
\label{sec:relatedwork}
%=======================================================================================
%====================================================
% Cohesion Metrics
%====================================================
There is much work on cohesion metrics.
Chidamber and Kemerer presented Lack of Cohesion Metric (LCOM1)
for measuring the number of the method pairs that reference no common attributes~\cite{Chidamber1991}.
The higher the number of methods pairs, the lower cohesion. 
Revising LCOM1, they presented LCOM2 to measure class cohesion by subtracting 
the number of the method pairs that share attributes from LCOM1~\cite{Chidamber1994}.
Li and Henry~\cite{Wei1993} redefine the concept of LCOM by defining sets of methods
that share an attribute. A method that shares an attribute with any method in a set
becomes a member of the set. To this end, the resulting sets are completely disjoint
and the number of the resulting sets indicates the cohesiveness of the class. 
That is, the higher the number of sets, the lower cohesiveness.   
Hitz and Montazeri~\cite{Hitz1995} represent LCOM by Li and Henry 
using undirected graphs where a node represents a method and an edge
represents attribute sharing by the paired methods.
Cohesiveness is then measured by the number of resulting graphs, which 
is known as LCOM3.
They also propose LCOM4 to take into account indirect reference to attributes. 
An edge is established between a method having a direct reference to an attribute
and a method invoking the directly referencing method.
Hendersen-Sellers~\cite{Henderson1995} proposes LCOM5 
which measures cohesion based on the number of referenced attributes.
The higher cohesion, the larger the number of referenced variables. 
Bieman and Kang~\cite{Bieman1995} proposed Tight Class Cohesion (TCC) 
and Loose Class Cohesion (LCC) based on direct and indirect 
connectivity among methods.
Marcus {\em et al}~\cite{Marcus2005} present Conceptual Cohesion of Classes (C3)
to measure method similarity based on textual coherence of in-line comments
and identifier names in source code.
Briand {\em et al.}~\cite{Briand1994} present Ratio of Cohesive Interaction (RCI)
which measures cohesiveness of a class as a ratio of the number of current interactions 
between method-to-data and data-to-data over the total number of possible interactions. 

%====================================================
% God class decomposition
%====================================================
Several researchers use a property-based approach for decomposing God classes
(e.g., see \cite{Simon2001,Fokaefs2009,Joshi2009,Cassell2009}).
The general approach is that the methods
of a God class are measured for their similarity based on method-attribute
relationships (e.g., methods sharing an attribute) and
method-method relationships (e.g., method calls) using
metrics (e.g., jaccard similarity metric \cite{SNEATH1973}).
The resulting similarity is then used as a basis for decomposing the God class.
Specifically, Simon {\em et al.}~\cite{Simon2001} measure similarity of
methods based on the distance of attribute use and method call.
Distance is short if one is dedicated to another (e.g.,
an attribute is used only in one method).
Extending the work by Simon {\em et al.}, Fokaefs and Tsantalis~\cite{Fokaefs2009}
use a clustering algorithm for decomposing properties of a God class.
Cassell {\em et al.}~\cite{Cassell2009} use
call graphs for presenting the relationship of methods and attributes
and employee the Girvan-Newman betweeness clustering algorithm~\cite{Girvan2002} for decomposing a Large class.
Extending the property-based approach, Bavota {\em et al.}~\cite{Bavota2011397}
make use of identifier names and in-line comments to measure 
similarity of methods using the LSI algorithm~\cite{Deerwester1990}, a technique for
measuring similarity of documents in the area of information retrieval.
The resulting similarity is represented in a weighted graph where
a node represents a method and an edge represents a pairwise relation of methods.
The MaxFlow-MinCut algorithm \cite{Cormen2001} is used to decompose
the similarity graph. Their work assumes that there exist ample
in-line comments and a naming convention for identifiers instilling
the intended context into the name, which is not always the case.

%====================================================
% God class detection
%====================================================
There exists some work on detecting God classes. 
Chatzigeorgiou {\em et al.}~\cite{Chatzigeorgiou2004} present
a design-based approach for identifying God classes.
They use collaboration diagrams to identify objects having
significant interactions by observing the number of links
between objects. Objects that have high fan-in and fan-out
are candidates of God classes.
However, that is not always the case, for example facade 
and proxy classes often have high fan-in and fan-out, but 
have a single responsibility of delegation. 
Joshi and Joshi~\cite{Joshi2009} present a lattice-based approach for identifying
less cohesive classes. A lattice captures attribute references in methods.
They propose seven types of lattices of which five types are cohesive and the other two
are less cohesive. A lattice conforming to the less cohesive types is advised to
be decomposed.
Marinescu~\cite{Marinescu2004} proposes metric-based rules for identifying God 
classes. They observe common symptoms of God classes such as
high complexity, low cohesiveness, and frequent access to data in other classes.
These symptoms are detected using  
Weighted Method Count (WMC)~\cite{Chidamber1994}, Tight Class Cohesion (TCC)~\cite{Bieman1995},
and Access To Foreign Data (ATFD)~\cite{Marinescu2001}.
Daniel {\em et al.}~\cite{Rapu2004} extend the work by Marinescu using historical data.
Classes are classified by frequency of change and the degree of change in size
observed in history. The higher frequency and degree of change, 
the more likelihood of being God classes.

%====================================================
% Summary
%====================================================
In summary, the existing work on detecting God classes and 
decomposing responsibilities heavily relies on intra-class information 
(e.g., attribute-method relationships, internal method dependencies, in-line comments).
However, object-oriented systems are collaborative by nature and it is
hard to derive precise semantics of methods without considering
class interactions. In this work, we use both intra-class information
and inter-class information with more emphasis on the latter.
Unlike the existing work, the presented approach can be used at both the design level and the code level.
At the design level, the approach can be used for class diagrams 
and sequence diagrams, which enables to detect God classes
early in development phase.

%=======================================================================================
\section{Overview of Approach}
\label{sec:overview}
\begin{figure}[!htb]
\centering
\scalebox{1.6}{\includegraphics{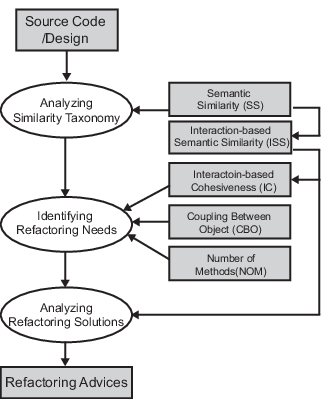}}
\caption{Process Overview}
\label{fig:process}
\end{figure}

%=======================================================================================
In this work, we view a class having a purpose for its existence.
In the view, we define a responsibility
as a set of methods to achieve the intended purpose of the class.
Given that, the approach aims at detecting God classes and
decomposing their responsibilities to be a single responsibility per class.
Figure~\ref{fig:process} shows an overview of the approach.
In the approach, God classes are detected based on pairwise semantic analysis of methods
using Resnik's taxonomy~\cite{Resnik1995}.
In the taxonomy, relative similarity for every pair of methods
is measured based on the architectural structure of the system using the Semantic Similarity (SS) metric.
The resulting similarity captures the structural distance between the paired methods
which we use as a base for measuring the semantic similarity of the methods.
The closer in distance, the more similar in semantics. The resulting similarity is then
refined by considering class relationships which are not taken into account
in the structural taxonomy. The refined similarity is then further refined
by considering internal method call dependencies within the same class.

A God class is detected using a set of metrics including Interaction-based Cohesiveness (IC),
Coupling Between Object (CBO)~\cite{Chidamber1994}, and Number Of Methods (NOM)~\cite{Bansiya2002}.
IC measures the cohesiveness of a class based on similarity of the methods that
interact with the class. Method interactions are measured using Interaction-based Semantic Similarity (ISS)
based on the structural taxonomy.
CBO measures the coupling of a class based on the interactions of the class with other classes, while
NOM measures the number of methods defined in a class.
Detected God classes are analyzed for their responsibilities using
a threshold determined by the average and standard deviation of ISS.
The resulting analysis advises a solution for decomposition of responsibilities.
We use complete weighted graphs to represent the solution.

%=======================================================================================
\section{Semantic Similarity}
\label{sec:csm}
\begin{figure}[!htb]
\begin{center}
\scalebox{1.55}{\includegraphics{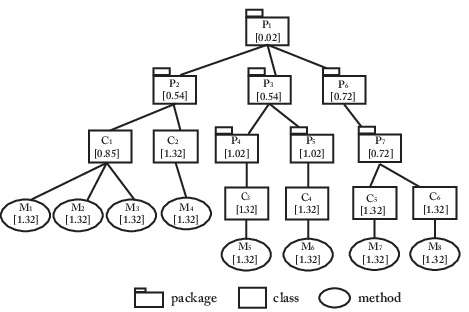}}
\end{center}
\caption{A Semantic Taxonomy of Entities}
\label{fig:TRS}
\end{figure}

%=======================================================================================
In this section, we describe analyzing semantic similarity of
methods by adopting Resnik's taxonomy~\cite{Resnik1995}.
Figure~\ref{fig:TRS} shows an example of the taxonomy
capturing the structure of a system in a tree
where leafs represent methods and non-leafs represent either classes or (sub)packages.
For example, in the figure, methods M1, M2, and M3 are defined in class C1
and classes C1 and C2 belong to package P2 which is a sub-package of P1.
Each node in the tree has its relative distance to other entities.
The distance is measured using the following metrics~\cite{cover2001}:

\begin{equation*}
SS(e_i, e_j) = -log P(ls(e_i, e_j))
\end{equation*}

where $ls(e_i, e_j)$ is the lowest superordinate of $e_i$ and $e_j$.

\begin{equation*}
P(e) = \frac{|se(e)|}{N}
\end{equation*}

where $se(e)$ is the set of sub-entities of $e$ and $N$ is the total number of nodes.

In Figure~\ref{fig:TRS}, the relative distance of $M_1$ and $M_4$ is measured 0.54 by
$SS(M_1, M_4) = -log P(ls(M_1, M_4))$ where $ls(M_1, M_4)$ is $P_2$.
$P(P_2)$ is $\frac{|se(P_2)|}{21}$ where $se(P_2)$ = \{$C_1$, $C_2$, $M_1$, $M_2$, $M_3$, $M_4$\}.
Thus, $P(P_2)$ = $\frac{6}{21}$ = 0.29 and $SS(M_1, M_4)$ = $-log 0.29$ = 0.54.
We use the distance as the semantic similarity of $M_1$ and $M_4$.
Similarly, the distance of $M_1$ and $M_5$ is measured 0.02.
The distances are interpreted that $M_1$ is more similar to $M_4$ in semantics than to $M_5$
since $M_1$ and $M_4$ belong to the same sub-package.
In the tree, leafs have the maximum similarity since their similarity is measured to themselves.
Table~\ref{table:CSM} shows the results of the taxonomy in matrix.
One may consider class libraries in the taxonomy for better results
if the results outweigh the overhead.

\begin{table}[!htb]
\centering
\caption{Similarity Matrix}
\scalebox{0.3}{\includegraphics{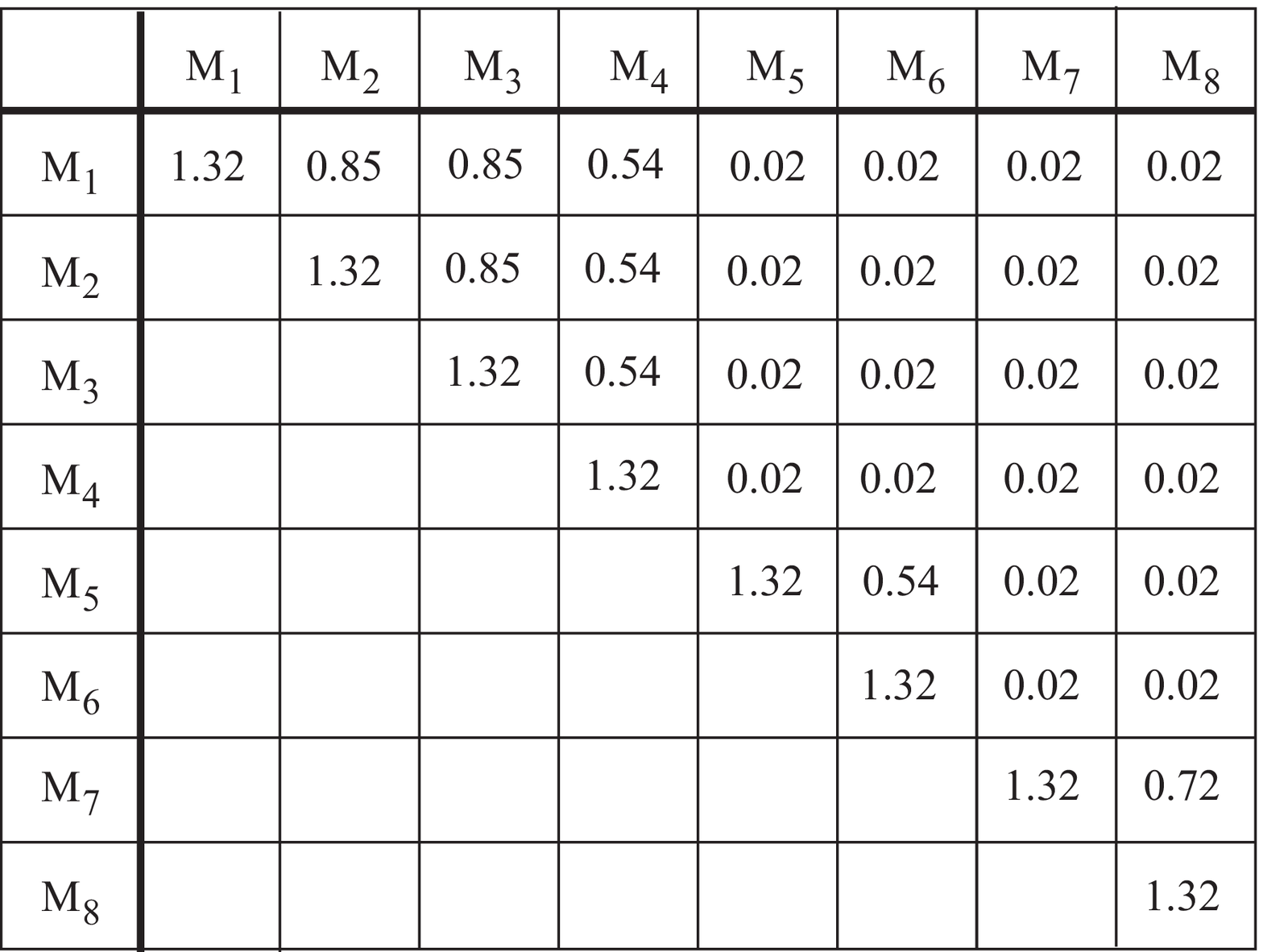}}
\label{table:CSM}
\end{table}

%================================================
\subsection{Refining Using Class Relationships}
\label{sec:refine1}
%================================================
The similarities in Table~\ref{table:CSM} consider only the structural
relationships of entities. We refine the similarities by considering
types of class relationships which are not captured in the taxonomy.
We consider 1) inner relationships, 2) generalizations, 3) aggregations,
4) associations and dependencies, which are in the order of high to low in weight.
Inner relationships are weighted 1.5, which is the highest, as the inner class
and the outer class have the full access each other.
Generalizations are weighted the second 1.4, since child
classes can inherit the properties of the parent class, but not all.
Aggregations are stronger than associations and dependencies due to
the whole-and-part constraint and weighted 1.3.
Associations and dependencies are weighted same 1.2 as they can be
interchangeably used, although dependencies are a little more limited in use.
The range is weights is determined in consideration of the relative
influence of class relationships to method similarity.
Considering these types with different weights refines the similarities
from the structural taxonomy.

\begin{figure}[!htb]
\begin{center}
\scalebox{0.8}{\includegraphics{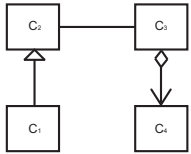}}
\end{center}
\caption{Class Relationships}
\label{fig:weighing-1}
\end{figure}

Table~\ref{table:sm_class} shows the refined similarities for the
relationships given in Figure~\ref{fig:weighing-1}. For instance, 
the similarity of methods $M_1$ and $M_3$ in Table~\ref{table:CSM} is 0.54
and it is refined to 0.76 (0.54 $\times$ 1.4) by considering the generalization in Figure~\ref{fig:weighing-1}.
Refined similarities are shown in bold in the table.  
Note that the refined similarity might be greater than
the maximum similarity (1.32) in Table~\ref{table:CSM}, which
is conceptually not valid (as nothing can be more similar than itself).
To remedy this, the minimum value that makes the maximum similarity
greater than the refined value in multiplication is used.
This value is referred to as {\em tapping constant}.

\begin{table}[!htb]
\centering
\caption{Refined Similarities with Weighed Class Relationships}
\scalebox{0.3}{\includegraphics{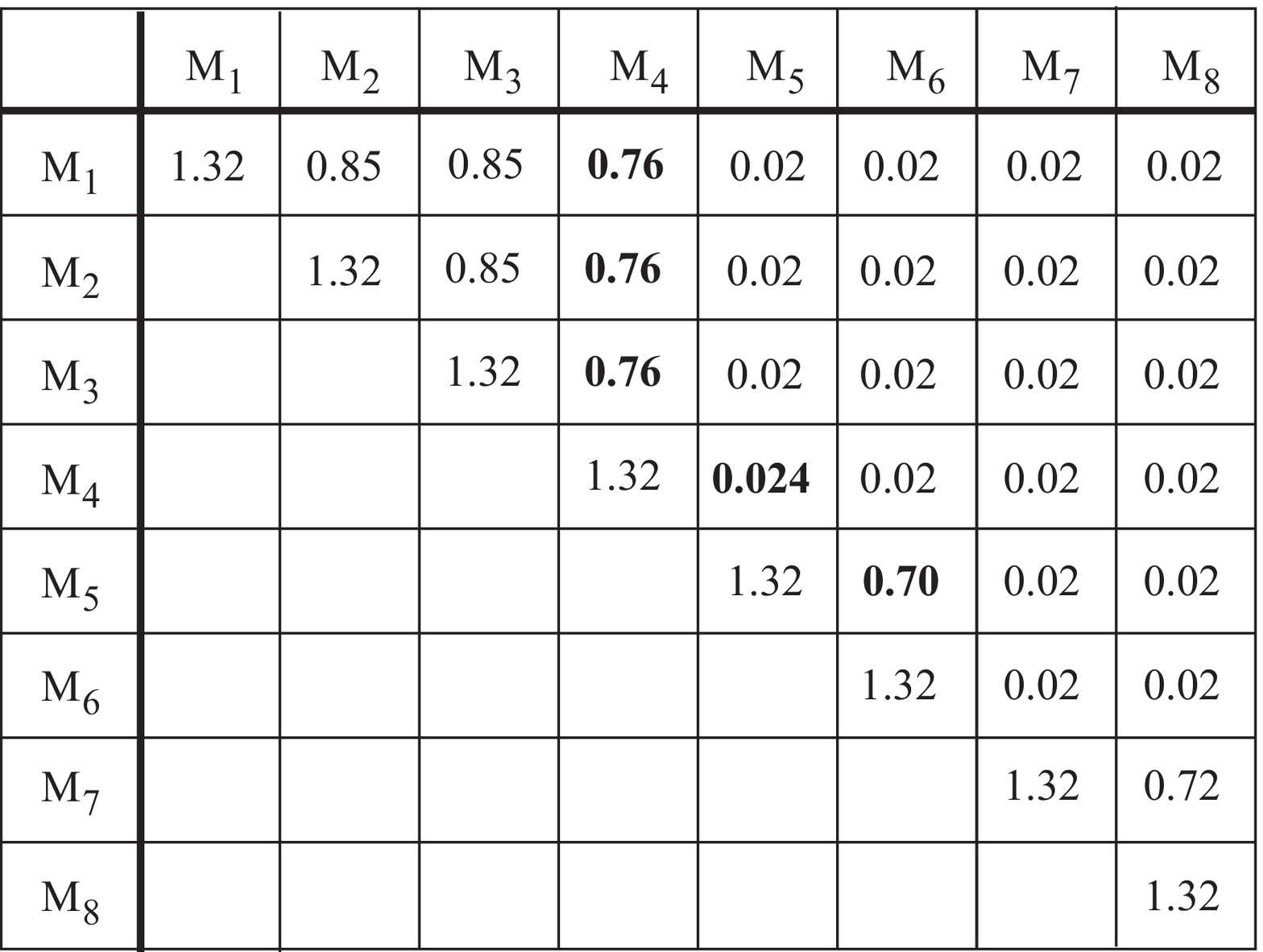}}
\label{table:sm_class}
\end{table}

%================================================
\subsection{Refining Using Method Call Dependencies}
\label{sec:refine2}
%================================================
We further refine the similarities resulting from Subsection~\ref{sec:refine1}
by considering method call dependencies within the same class.
Having a call dependency between methods belonging to the same class
shows a great intimacy and their similarity is doubled.
Method call dependencies are weighed higher than class relationships
as method dependencies are more influential to the semantic similarity of methods
than class relationships. 
Suppose method $M_1$ has a call dependency on method $M_2$.
Given this, the similarities in Table~\ref{table:sm_class} are
refined as shown in Table~\ref{table:sm_method} where
the similarity of $M_1$ and $M_2$ is refined to 1.7 from 0.85.
A tapping constant can be used if the refined value
becomes greater than the maximum similarity.

\begin{table}[!htb]
\centering
\caption{Refined Similarities with Weighed Call Dependency within the Same Class}
\scalebox{0.3}{\includegraphics{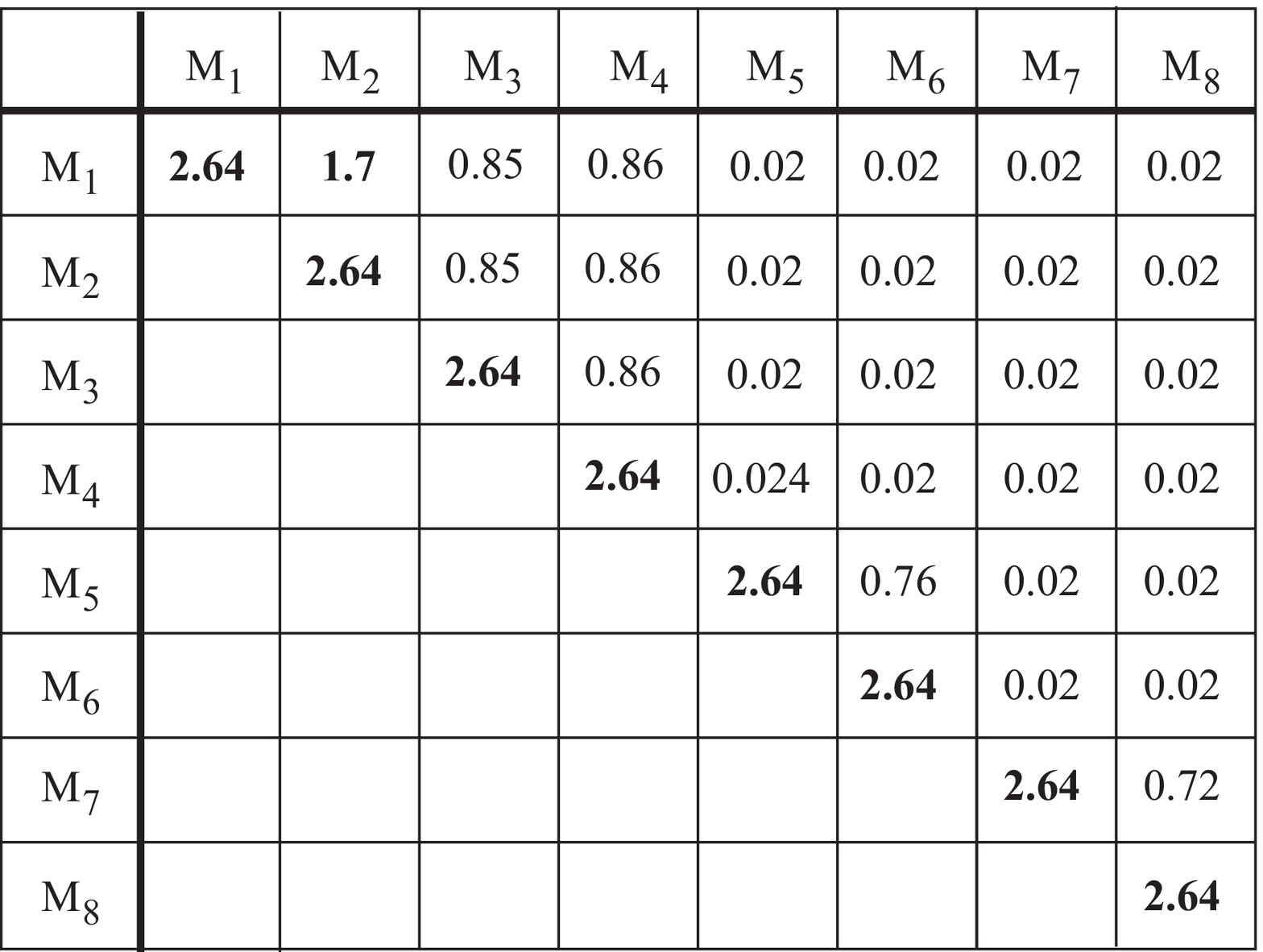}}
\label{table:sm_method}
\end{table}

%=======================================================================================
\section{Extracting responsibilities of God Class}
\label{sec:badsmell}
\begin{figure}[!htb]
\centering
\scalebox{1.5}{\includegraphics{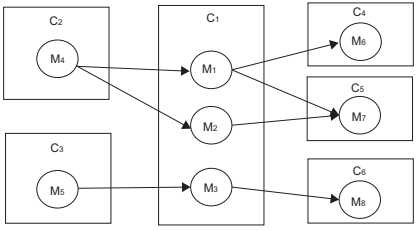}}
\caption{Fan-In and Fan-Out of Methods}
\label{fig:exampleICM}
\end{figure}

%=======================================================================================
The similarities resulting from Section~\ref{sec:csm} capture only the similarity
of methods across classes and do not fully capture the similarities of the methods
in the same class. This is because the base taxonomy used in Section~\ref{sec:csm} is built upon the structure of entities
which does not carry any internal information of a class.
For this reason, the same similarity is shown in Figure~\ref{fig:TRS} for the methods
in the same class (e.g., methods $M_1$, $M_2$, and $M_3$ in class $C_1$ have
the same similarity 0.85).
The refinement in Subsection~\ref{sec:refine2} measures limited
similarity of internal methods by considering internal method dependencies.
For the full extent of measuring the similarity of the same class methods,
we make use of method interactions.
That is, the similarity of the same class methods are measured
``indirectly'' through the similarity of their interacting methods in other classes,
which is referred to as Interaction-based Semantic Similarity (ISS).
ISS of methods $m_i$ and $m_j$ is measured as follows:

\begin{align*}
ISS(m_i, m_j)=&mss(F_{in}(m_i), F_{in}(m_j)) \\
             &+mss(F_{out}(m_i), F_{out}(m_j)) \\
             &+SS(m_i, m_j) \\
             \\
mss(ms_i, ms_j) =& \frac{\sum_{m_i \in ms_i}\sum_{m_j \in ms_j} SS(m_i, m_j)}{|ms_i| \times |ms_j|}
\end{align*}

where $F_{in}(m)$ is the fan-in function of method $m$ returning the set of invoking
methods for $m$ and $F_{out}(m)$ is the fan-out function returning the set of invoked
methods and SS($m_i$, $m_j$) is the similarity of methods $m_i$ and $m_j$ found
in Table~\ref{table:sm_method}.
For example, suppose that we measure the similarity of $M_1$ and $M_2$ in Figure~\ref{fig:exampleICM}.
In the figure, the fan-in of $M_1$ and $M_2$ is \{$M_4$\},
the fan-out of $M_1$ is \{$M_6$, $M_7$\}, and the fan-out of $M_2$ is \{$M_7$\}.
Based on Table~\ref{table:sm_method}, $mss$(\{$M_4$\}, \{$M_4$\}) = 2.64,
$mss$(\{$M_6$, $M_7$\}, \{$M_7$\}) = 1.33 and $SS(M_1, M_2)$ = 1.7, which
results in ISS($M_1$, $M_2$) = 5.67. The similarity of other pairs
can be computed (ISS($M_1$, $M_3$) = 1.24 and ISS($M_2$, $M_3$) = 1.59).

%================================================
\subsection{Detecting God Classes}
\label{sec:dgc}
\begin{figure}[!htb]
\centering
\scalebox{2.5}{\includegraphics{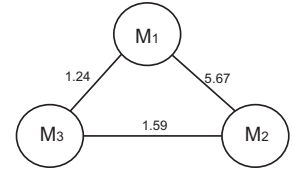}}
\caption{Example of complete weighted graph}
\label{fig:exampleofweightedgraph}
\end{figure}

%================================================
God classes are often large in size and have high coupling with other
classes and low cohesion with respect to the similarity of interacting
methods~\cite{Brown1998,Fowler1999}.
Given this observation, we identify God classes using a set of metrics
including Interaction-based Cohesiveness (IC),
Number of Methods (NOM)~\cite{Bansiya2002}, and
Coupling Between Objects (CBO)~\cite{Chidamber1994}.
IC measures the cohesiveness of a class based on the similarity of the methods that
interact with the class as follows:

\begin{displaymath}
IC(c) = \frac{\sum_{m_i\in M}\sum_{m_j\in M}ISS(m_i, m_j)}{|M| \times (|M|-1)}
\end{displaymath}

where $M$ is the set of the methods in class $c$ and ${m_i \neq m_j}$.
For instance, IC of the class $C_1$ in Figure~\ref{fig:exampleICM} is measured 2.18.
CBO measures the coupling of a class based on the interactions of the class with other classes, while
NOM measures the number of methods defined in a class.
Using IC, CBO, and NOM, we define the following rule for detecting God classes:

\begin{align*}
GC(S) = & \{c \in C |  (NOM(c) > 3rd Quartile (S))\\
              \wedge &(CBO(c) > 3rd Quartile(S))\\
              \wedge &(IC(c) < 3rd Quartile(S))\}
\end{align*}

where $C$ is the set of classes defined in system $S$.
In the rule, IC, NOM, and CBO are set to the 3rd quartile as a default.
We use box plots to represent statistical filtering.
A detected god class is represented using a complete weighted graph
where a node represents a method and the weight on each edge represents
the semantic similarity of the paired methods linked by the edge.
Figure~\ref{fig:exampleofweightedgraph} shows an example graph
for the class $C_1$ in Figure \ref{fig:TRS}.

%================================================
\subsection{Decomposing Responsibilities}
\label{sec:ergc}
\begin{figure}[!htb]
\centering
\scalebox{2.5}{\includegraphics{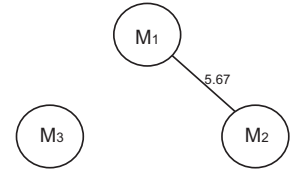}}
\caption{Splitting Responsibilities}
\label{fig:exampleofextracting}
\end{figure}

%================================================
God class graphs resulting from Subsection~\ref{sec:dgc}
are analyzed for decomposition of responsibilities using a threshold.
A threshold $\epsilon$ is determined as follows:

\begin{align*}
\epsilon &= [\mu - \sigma, \mu + \sigma] \\
\\
&if \ \mu - \sigma < MinWeight \ then \ \epsilon = MinWeight \\
&if \ \mu + \sigma > MaxWeight \ then \ \epsilon = MaxWeight
\end{align*}

where $\mu$ is the average of weights and $\sigma$ is the standard deviation
of weights. The threshold guarantees the edge of two nodes having
the weight (similarity) lower than the threshold to be removed,
which splits the God class graph into sub-graphs, each
capturing a single responsibility.
Figure \ref{fig:exampleofextracting} shows two sub-graphs split
from the God class graph Figure \ref{fig:exampleofweightedgraph}
by a threshold ranging from 0.37 to 5.30.

%=======================================================================================
\section{Case Study: JMeter}
\label{sec:casestudy}
\begin{figure}[!htb]
\centering
\scalebox{1.6}{\includegraphics{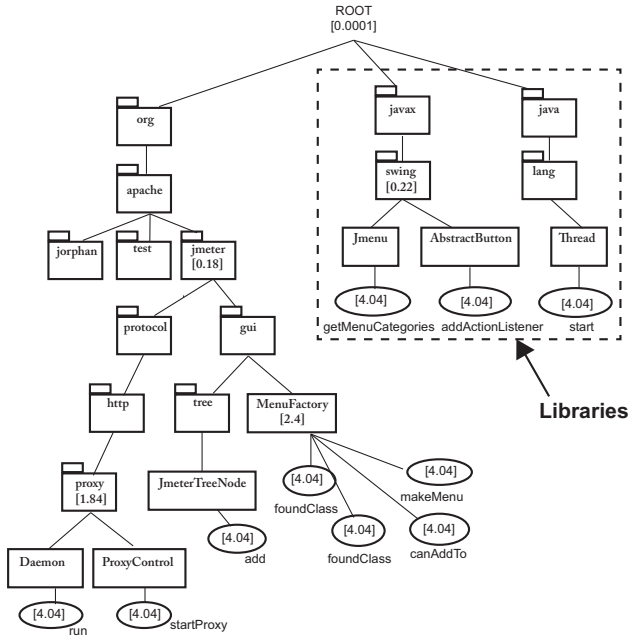}}
\caption{Structural Taxonomy of JMeter v2.5.1}
\label{fig:case_tr}
\end{figure}

%=======================================================================================
We use JMeter v.2.5.1~\cite{JMeter}, a Java-based open source for load testing and measuring
performance of a server, to evaluate the presented approach. The version used in this study
involves 405 packages, 1,623 classes, 9,005 methods, and 30 libraries with 2.5 years of maintenance.
The size of the application is 145 KLOC and the average NOM and the average CBO are 8.5 and 11.1, respectively.
In applying the approach, the structural taxonomy involves 11,033 entities in total including
classes, methods, and packages, which results in a 9,005 $\times$ 9,005 similarity matrix.
Figure~\ref{fig:case_tr} shows partial results of the taxonomy and
a corresponding matrix is shown in Table~\ref{table:case_sm}.
In this study, we also consider libraries in similarity analysis.
The dashed box in Figure~\ref{fig:case_tr} shows a subset of the considered libraries.
The great deviation between the minimum value and the maximum value in Table~\ref{table:case_sm}
is a hint of the large number of entities used in this study.

\begin{table}[!htb]
\centering
\caption{Semantic Similarities of JMeter v2.5.1}
\scalebox{0.3}{\includegraphics{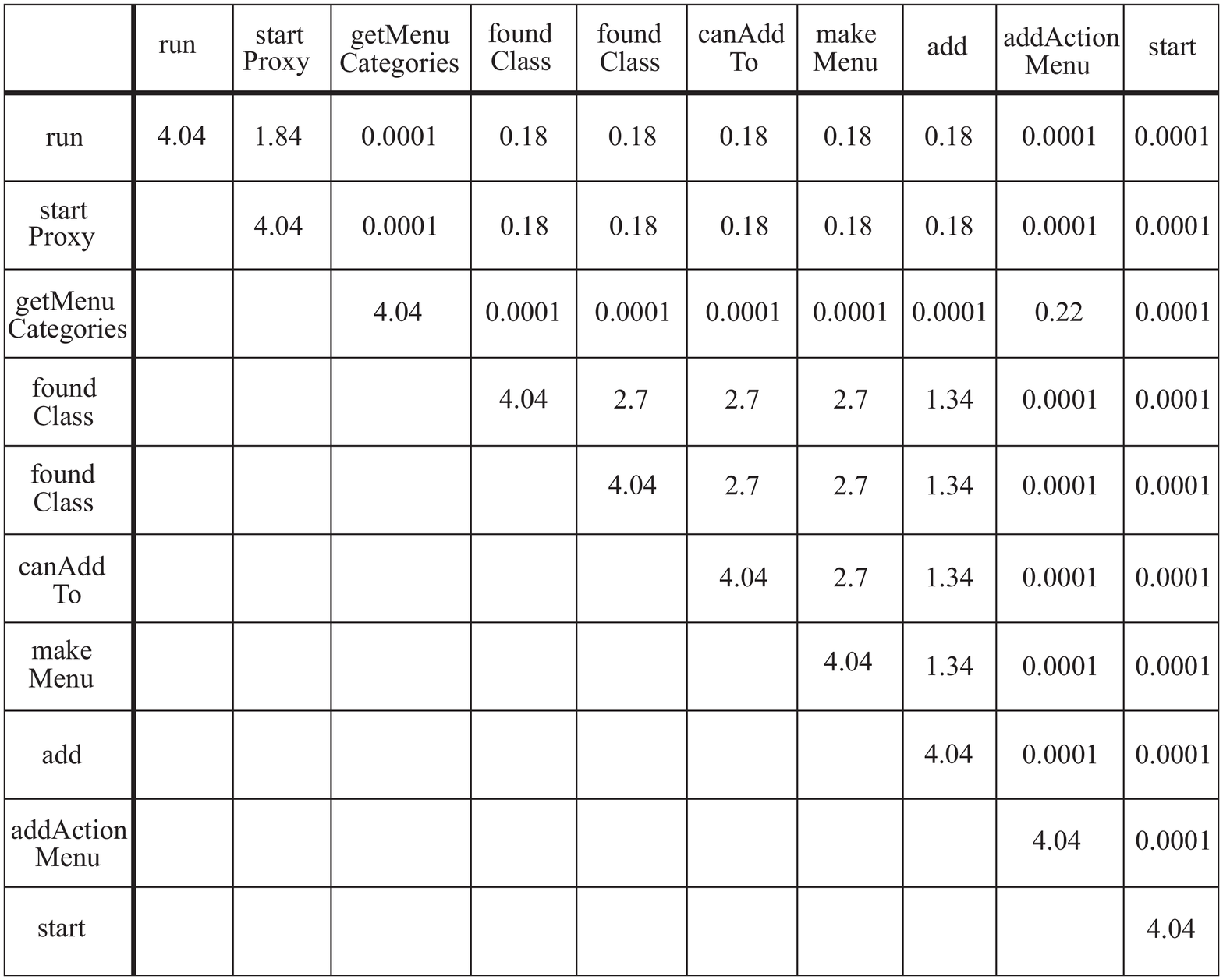}}
\label{table:case_sm}
\end{table}

The similarities in Table~\ref{table:case_sm} are refined in consideration
of class relationships. There are 176 inner class relationships,
187 generalizations, 655 associations/dependencies
found. Applying the weights for class relationships in Section~\ref{sec:csm},
the similarities are refined to Table~\ref{table:case_sm-1}.

\begin{table}[!htb]
\centering
\caption{Refined Similarities of JMeter v2.5.1 with Class Relationships}
\scalebox{0.3}{\includegraphics{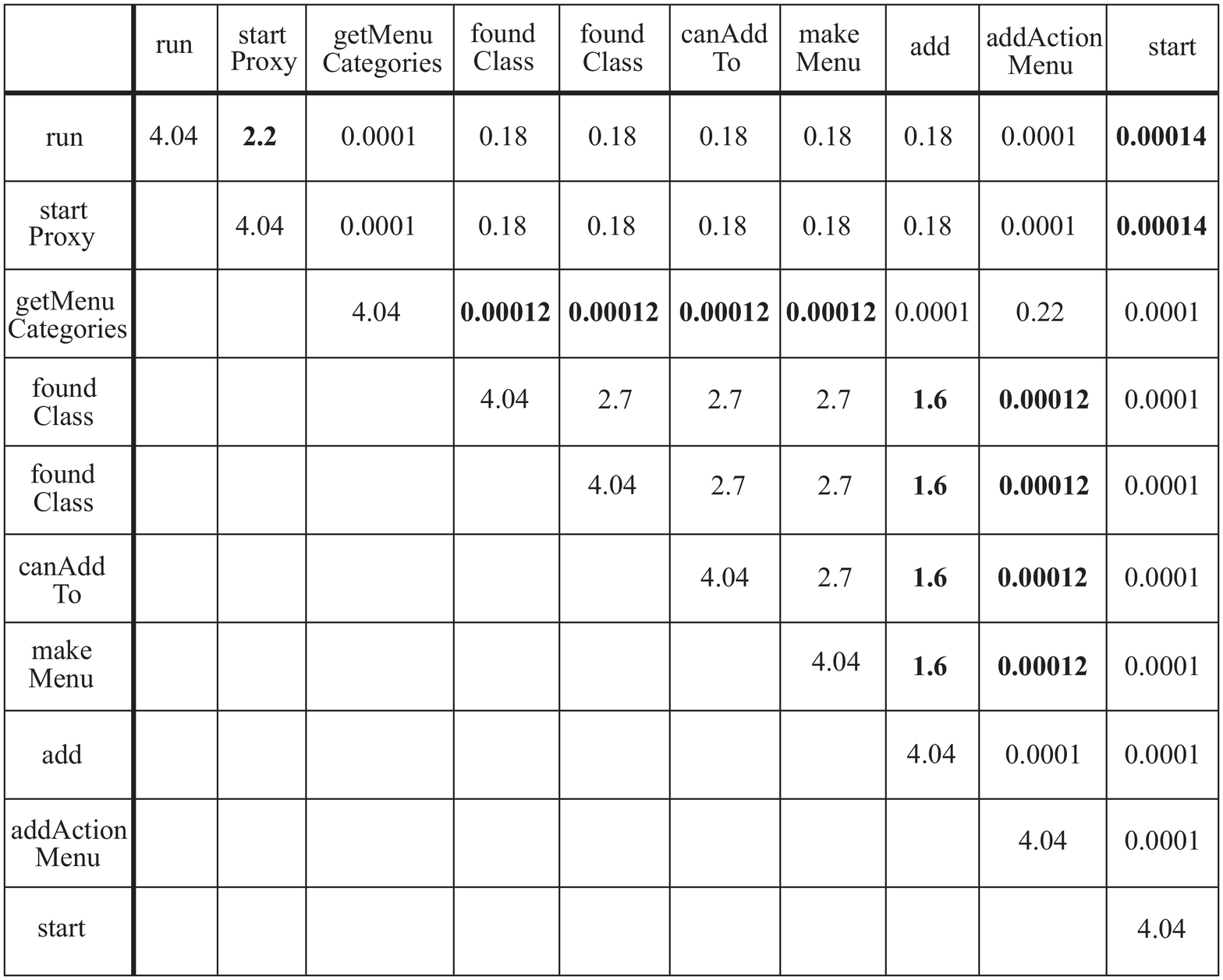}}
\label{table:case_sm-1}
\end{table}

The similarities in Table~\ref{table:case_sm-1} are further refined
in consideration of method call dependencies which involve
4,066 dependencies. Table~\ref{table:case_sm-2} shows the refined similarities.

\begin{table}[!htb]
\centering
\caption{Refined Similarities of JMeter v2.5.1 with Method Call Dependencies}
\scalebox{0.3}{\includegraphics{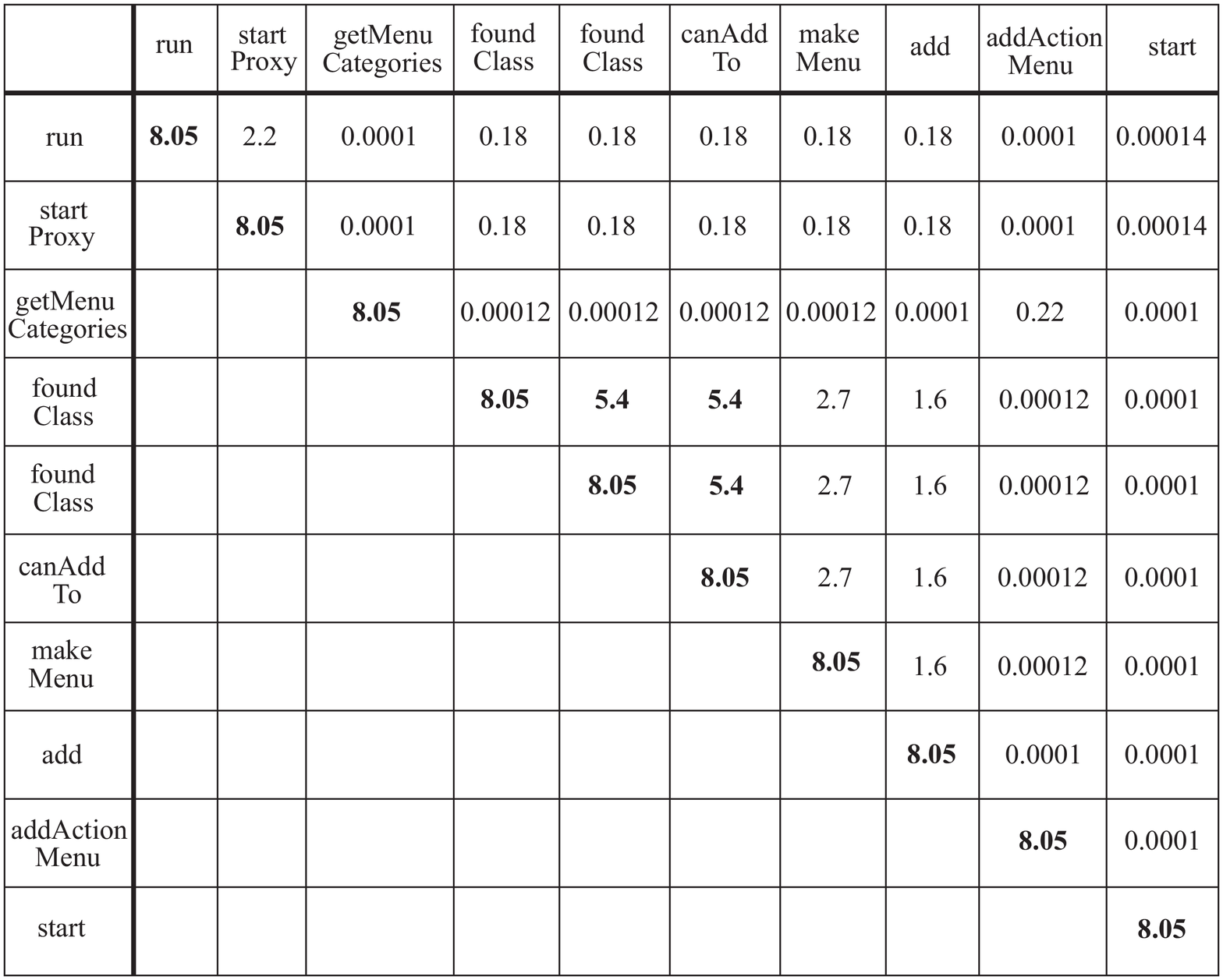}}
\label{table:case_sm-2}
\end{table}

\begin{figure}[!htb]
\centering
\scalebox{1.6}{\includegraphics{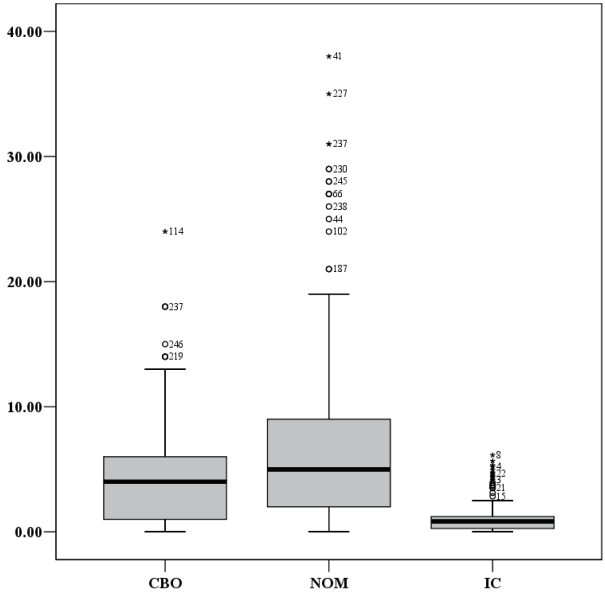}}
\caption{Measures of CBO, NOM, and IC}
\label{fig:casestudy-1}
\end{figure}

Based on the resulting similarities in Table~\ref{table:case_sm-2},
the 3rd quartile of CBO, NOM, and IC is measured
9.0, 6.0 and 1.22, respectively, which leads to the
following detecting rule:

\begin{align*}
GC(JMeter v2.5.1) = & \{c \in C |  (NOM(c) > 9.0)\\
              \wedge &(CBO(c) > 6.0)\\
              \wedge &(IC(c) < 1.22)\}
\end{align*}

where $C$ is the set of the classes in JMeter v2.5.1.
Figure \ref{fig:casestudy-1} shows box plots for CBO, NOM, and IC.

By applying the rule, six classes out of 1,623 classes are detected
as candidate God classes. Table~\ref{table:godclasscandidate} shows
the list of the detected classes.

\begin{table}[!htb]
  \centering
  \caption{Candidate God Classes}
  \scalebox{0.5}{\includegraphics{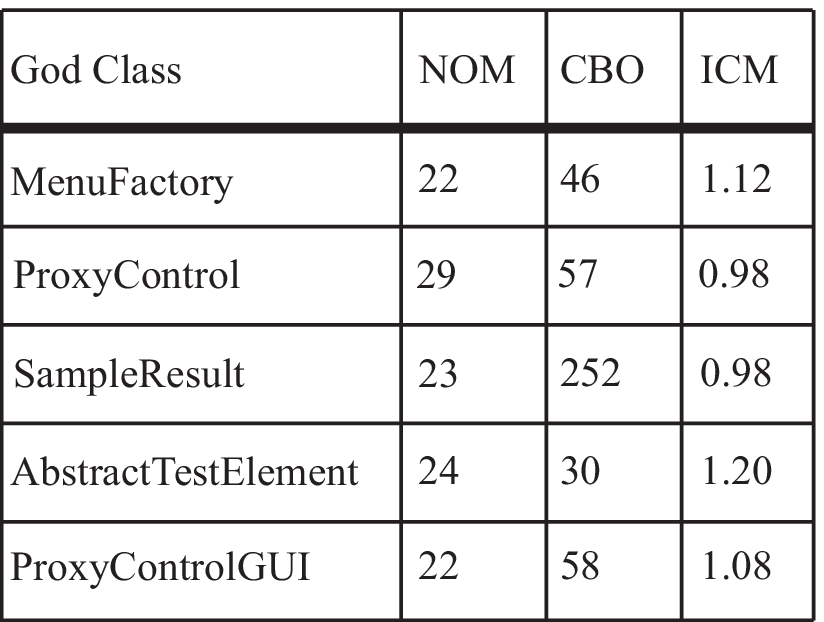}}
  \label{table:godclasscandidate}
\end{table}

\begin{figure*}
\centering
\scalebox{1.9}{\includegraphics{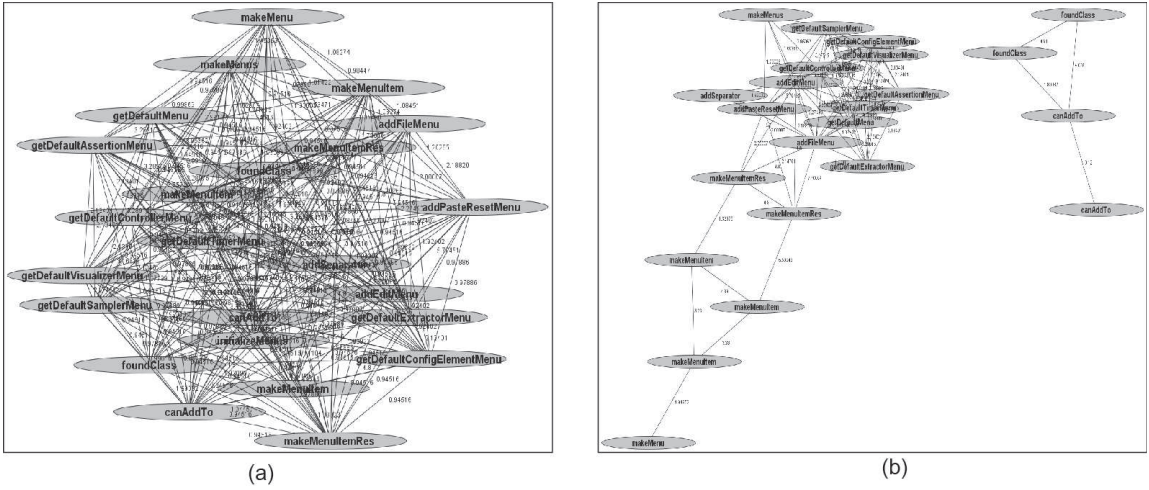}}
\caption{Decomposing Responsibilities of the $MenuFactory$ Class}
\label{fig:splitresult}
\end{figure*}

For each candidate class, a complete weighted graph is
built. Figure~\ref{fig:splitresult}(a) shows the graph for
the $MenuFactory$ class. The graph involves 22 nodes
representing the methods defined in the class and 231 edges connecting the nodes.
The threshold for the class ranges from 0.95 to 2.09
and we experimented a threshold ranging from 1 to 2.1 incremented by 0.1.
No edge is removed with threshold 1.0 causing no split and
217 edges are removed with threshold 2.1 resulting in three sub-graphs.
In a manually verification, threshold 1.5 produces
the best result, which decomposes into two sub-graphs with
157 edges removed, each sub-graph representing a single responsibility.
One sub-graph involves 18 methods and the other involves 4 methods.
Figure~\ref{fig:splitresult}(b) shows the resulting decomposition.
Table \ref{table:thinterval} shows the threshold ranges used in
experiments for each of the candidate God classes.

\begin{table}[!htb]
  \centering
  \caption{Thresholds and ISS Statistics}
  \scalebox{0.5}{\includegraphics{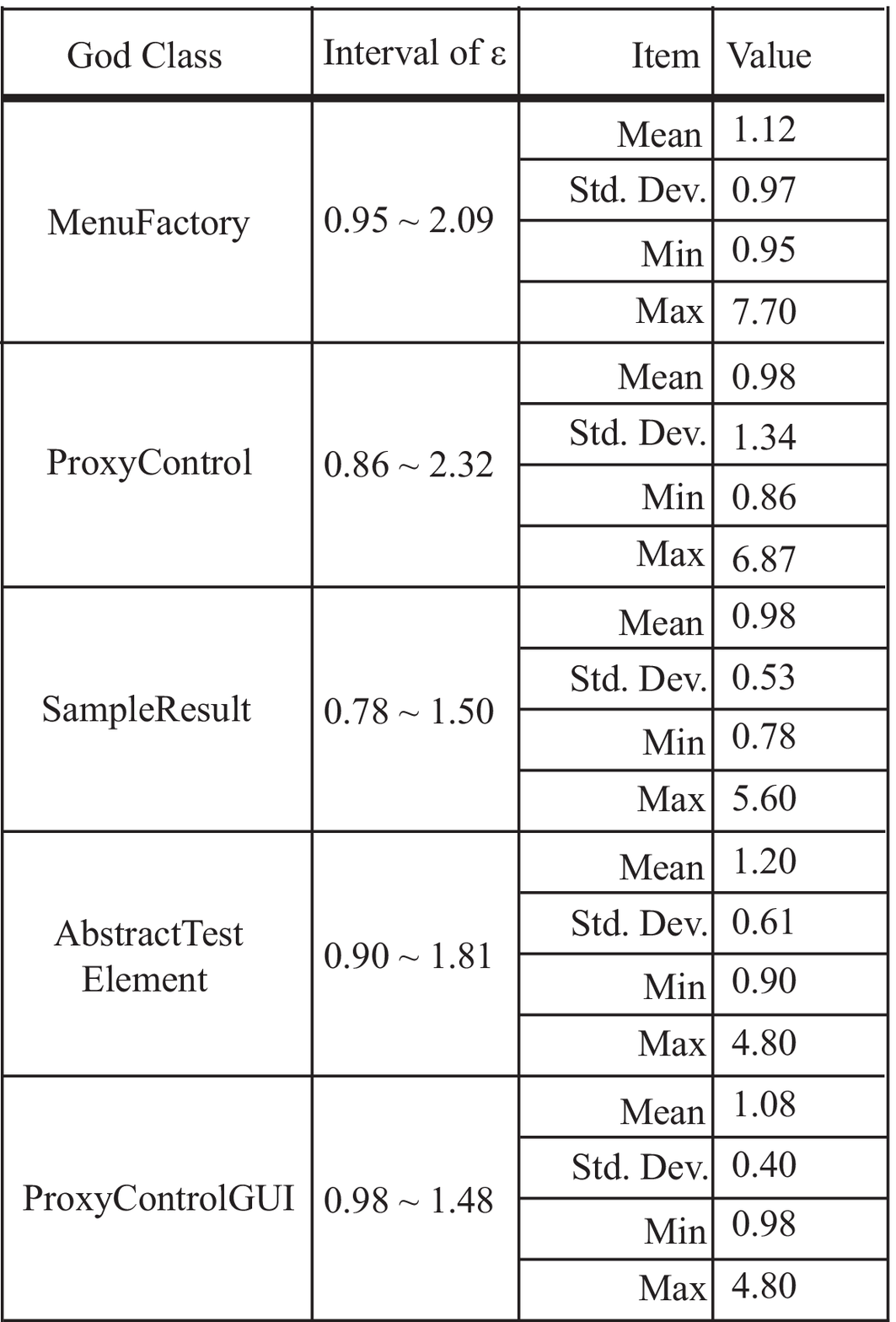}}
  \label{table:thinterval}
\end{table}

%================================================
\subsection{Results Analysis}
\label{sec:car}
\begin{figure}[!htb]
\centering
\scalebox{1.2}{\includegraphics{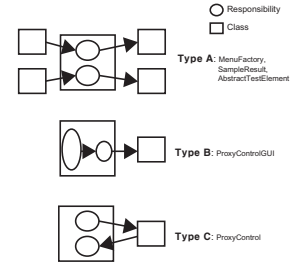}}
\caption{Three Types of Detected God Classes}
\label{fig:types}
\end{figure}

%================================================
From the case study, we observe three types of God classes
shown in Figure \ref{fig:types}.
Type A has two heterogeneous responsibilities, each having an independent
set of fan-in and fan-out interactions, which should be put in a separate
class. This is an example of a obvious need for responsibility decomposition.
The \em MenuFactory\em , \em SampleResult \em, and \em AbstractTestElement \em classes belong to Type A.
The \em ManuFactory \em class has responsibilities of 1) creating menus and 2) controlling the drag and drop function.
While the drag and drop function supports menus, its controlling
responsibility is not directly related to menus.
The $SampleResult$ class involves responsibilities of 1) collecting and storing sample results
and 2) measuring the time taken to collect sample results.
The collecting and storing functions in the first responsibility is
quite heterogeneous to the measuring function in the
second responsibility.
The \em AbstractTestElement\em class has responsibilities of
1) configuring properties of tested elements and 2)configuring thread context.
The target objects concerned in the two responsibilities are
completely different types, and thus the responsibilities share no commonality.

Type B also involves two responsibilities. However, unlike Type A,
the responsibilities in Type B have a dependency. When the responsibilities
are split into two classes, the dependency is realized as an association
between the classes.
The $ProxyControlGUI$ class is in Type B.
As the name implies, the class involves two dependent responsibilities
including 1) creating and controlling GUI and 2) controlling proxies.
The proxy responsibility should be separated and put into the $ProxyControl$ class
which is associated with the $ProxyControlGUI$ class.
Type B is an example of the {\em Feature Envy} smell~\cite{Fowler1999}
which violates the principle of grouping behaviors by related data
and occurs when a method is more interested in being in another class than the current class.

Type C have also two responsibilities that have an indirect dependency via a class.
At the class level, the dependency appears as a bidirectional dependency (or association).
From an implementation view, a bidirectional dependency is costly
as two-way links should be manipulated for creating, accessing, and removing
objects, which would not have to be dealt if the responsibilities are separated.
The $ProxyControl$ class belongs to Type C.
The $ProxyControl$ class involves responsibilities of 1) starting and stopping proxies
and notifying start and stop of proxies to other objects and
2) receiving the resulting data from proxies and delegating the data to other objects.
The first responsibility depends on the $Proxy$ class, which in turn depends on
the second responsibility.
Such a bidirectional association is not a common practice and
is often a source of errors ~\cite{Fowler1999}.

%================================================
\subsection{Verifying Results}
\label{sec:verification}
%================================================
We verify the accuracy of the decomposition results by comparing them
with manual results produced by two software engineers
having 5-year and 2-year industry experience.
The engineers manually reviewed the detected classes and
their interacting classes for identifying responsibilities.
The results of the manual review are shown in Table~\ref{table:bestaccuracymeasure}.

\begin{table*}[!htb]
\centering
\caption{The best accuracy measure}
\scalebox{0.5}{\includegraphics{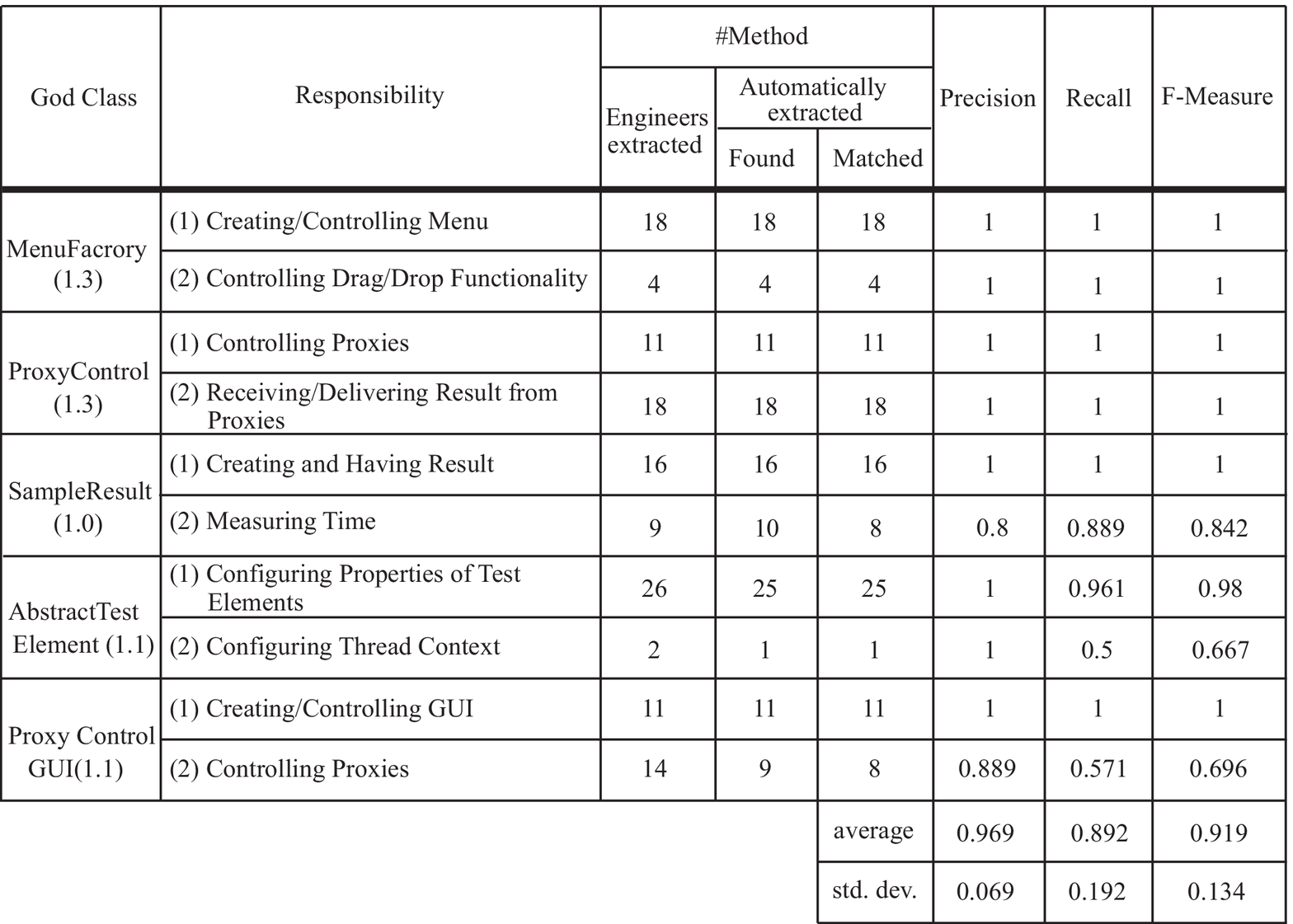}}
\label{table:bestaccuracymeasure}
\end{table*}

Let $R(c)=\{r_1, r_2, ... ,r_n\}$ be the set of manually identified
responsibilities of class $c$ where $r_i= \{m_1, m_2, ..., m_p\}$ is a set of methods.
Let $R'(c)=\{r_1', r_2', ... ,r_m'\}$ be the set of the responsibilities identified
by the presented approach where $r_i' = \{m_1', m_2', ..., m_q'\}$ is a set of methods.
For $r_i \in R$ and $r_j' \in R'$, the best matching responsibility in $R'$ is
$r_{best}' = argmax \ \frac{|\ r_j' \cap \ r_i \ |}{|\ r_j' \cup \ r_i \ |}$.
Given that, the accuracy of the presented approach can be measured using
the following precision, recall, and F-Measure:

\begin{itemize}
\item Precision: The number of correctly identified methods of a responsibility $r_i$
to its best matching responsibility $r_{best}'$ over the number of the identified methods of $r_i$.
\begin{displaymath}
Precision: \ P(r_i) = \frac{|\ r_i \ \cap \ r_{best}'\ |}{| \ r_{best}' \ |}
\end{displaymath}

\item Recall: The number of correctly identified methods of a responsibility $r_i$
to its best matching responsibility $r_{best}'$ over the number of the defining methods of $r_i$.
\begin{displaymath}
Recall: \ R(r_i) = \frac{|\ r_i \ \cap \ r_{best}'\ |}{| \ r_i \ |}
\end{displaymath}

\item F-Measure: A composite measure of $P(r_i)$ and $R(r_i)$ for responsibility $r_i$.

\begin{displaymath}
F-Measure: \ F(r_i) = \frac{2 \cdot P(r_i) \cdot  R(r_i)}{P(r_i) + R(r_i)}
\end{displaymath}
\begin{displaymath}
F(GC) = \frac{2 \cdot \frac{{\textstyle \sum_{r_i \in R} P(r_i) }}{|R|}  \cdot \frac{{\textstyle \sum_{r_i \in R} R(r_i) }}{|R|}}{\frac{{\textstyle \sum_{r_i \in R} P(r_i)}}{|R|}  + \frac{{\textstyle \sum_{r_i \in R} R(r_i) }}{|R|} }
\end{displaymath}
\end{itemize}

\begin{figure}[!htb]
\centering
\scalebox{1.2}{\includegraphics{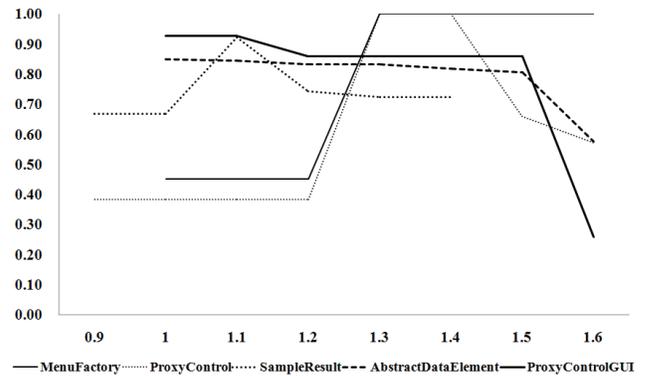}}
\caption{F-Measure for Detected God Classes}
\label{fig:accuracymeasure}
\end{figure}

Figure~\ref{fig:accuracymeasure} shows the results of F-Measure
for the detected God classes per change of threshold.
The results in the graph are capricious, which indicates a high deviation of ISS.
In fact, the likelihood of being split for a God class graph increases
as the deviation of edge weights increases. Therefore, the results
indicate a high likelihood of decomposition for the detected classes.
The graph also show high accuracy of the results in the threshold range of 1.3 and 1.5.

In Table~\ref{table:bestaccuracymeasure}, the average of the best F-Measures
for the detected God classes is 0.919 which is promising.
In particular, the responsibilities of the $MenuFactory$ and $ProxyControl$ class
are identified exactly the same as the manually found ones.
In this study, the accuracy of the detecting rule is not measured
as it is not feasible to identify the actually existing God classes
in JMeter due to subjectivity and the large number of classes (1,623).

%In the table,
%TODO discuss how association is generated differently from dependency in tool use.

%=======================================================================================
\section{Conclusion}
\label{sec:conclusion}
%=======================================================================================
In this paper, we have presented a semantic-based approach for detecting 
God classes and decomposing their responsibilities. 
The approach measures semantic similarity of methods using
inter and intra-interactions of methods and class relationships.
The resulting similarity is 
used as a basis for identifying God classes using the NOM, CBO, and IC metrics. 
Detected God classes are represented in a weighted graph
for responsibility analysis and decomposition.
Responsibilities are identified based on relative semantic similarity 
of defined methods in the God class to the similarity of their 
interacting methods in other classes.
Responsibilities are decomposed by a threshold 
determined by the average and standard deviation of ISS. 
We evaluate the approach using JMeter v2.5.1.

The presented approach does not require code details (e.g., attribute-method references),
and therefore can be used at both design level and code level. 
In this paper, we did not consider constructors, getters, and setters
since they are not captured at design level. 
At code level, however, constructors may be a subject of interest
in detecting God classes and decomposing responsibilities. 
Getters and setters at code level can help to improve the precision of similarity
if they are referenced in other methods, which basically captures
attribute-method references. 
Similarity precision can be further improved if libraries are 
considered in the taxonomy. However, it involves an overhead and 
is recommended only when the accuracy of the expected results outweigh the overhead.

%================================================================
\noindent{\bf\large Acknowledgements.}\\
%================================================================
\noindent This research was supported by the MKE(The Ministry of Knowledge Economy), Korea, under the ITRC(Information Technology Research Center) support program supervised by the NIPA. (National IT Industry Promotion Agency(NIPA-2011-(C1090-1131-0008)))

\bibliographystyle{unsrt}
\bibliography{zuna}
\end{document}